\documentclass[%        	Class options:
aps,%                   	American Physical Society
prd,%                   	Physical Review D
showpacs,%              	Displays PACS after abstract
twocolumn,
superscriptaddress,%    	Authors' addresses linked with superscripts
nofootinbib,%           	Does not treat footnotes as references
floatfix]%              	Fixes float errors
{revtex4-1}%              	REVTEX 4 Package used
\usepackage{graphicx,%  	Default Latex 2eps package for embedding figures
longtable,%             	Useful for long table
slashed,%						R-parity slash
hyperref,
bm,url,
braket,
xcolor,
xspace}
\usepackage[normalem]{ulem}

\begin{document}

%

%==================================================
%: Change
\title{Arbitrary mass Majorana neutrinos in neutrinoless double beta decay}
%\title{Sterile Neutrinos in Neutrinoless Double Beta Decay: An Update}
%
%-------------------------------------------------------------------------------------
%
%
\author{Amand Faessler}
\affiliation{Institute f\"ur Theoretische Physik der Univesit\"at
T\"ubingen, Auf der Morgenstelle 14, D-72076 T\"ubingen, Germany}
\author{Marcela Gonz\'alez}
\author{Sergey Kovalenko}
\affiliation{Universidad T\'{e}cnica Federico Santa Mar\'ia,
Centro-Cientifico-Tecnol\'{o}gico de Valpara\'iso,
Casilla 110-V, Valpara\'iso, Chile}
\author{F. \v{S}imkovic}
\affiliation{Bogoliubov Laboratory of Theoretical Physics, JINR 141980 Dubna,
Russia}
\affiliation{Department of Nuclear Physics and Biophysics, Comenius
University, Mlynsk\'{a} dolina F1, SK-842 48
Bratislava, Slovakia}
\affiliation{Czech Technical University in Prague, 128-00 Prague, Czech Republic}

\begin{abstract}%.......................................................................
We revisit the mechanism of neutrinoless double beta ($0\nu\beta\beta$) decay mediated by  
the exchange with the heavy Majorana neutrino N of arbitrary mass $m_{\rm N}$, slightly mixed  
$\sim U_{e\rm N}$ with the electron neutrino $\nu_{\rm e}$. By assuming the dominance 
of this mechanism, we update the well-known $0\nu\beta\beta$-decay 
exclusion plot in the \mbox{$m_{\rm N}-U_{e\rm N}$} plane taking into account recent progress in 
the calculation of nuclear matrix elements within quasiparticle random phase approximation 
and improved experimental bounds on the $0\nu\beta\beta$-decay half-life of ${^{76}}$Ge 
and ${^{136}}$Xe. We also consider the known formula approximating the $m_{\rm N}$
dependence of the $0\nu\beta\beta$-decay nuclear matrix element in a simple explicit form. 
We analyze its accuracy and specify the corresponding  parameters,  
allowing one to easily calculate the $0\nu\beta\beta$-decay half-life for arbitrary $m_{\rm N}$
for all the experimentally interesting isotopes without resorting to real nuclear structure calculations.  
%
%It is shown that the dependence of matrix element on the mass of heavy
%neutrino can be approximated by a simple analytic formula reproducing correctly asymptotics 
%for zero neutrino mass and extremely large mass of neutrino. The exclusion \mbox{$m_{\rm N}-U_{\rm eN}$} 
%plot is compared with similar constraints derived in the literature from other processes.  
%
 \end{abstract}%.......................................................................

\pacs{11.30.Fs, 14.60.Pq, 14.60.St, 23.40.-s, 23.40.Bw, 23.40.Hc}

\keywords{Sterile neutrino, Lepton number violation, Neutrinoless double beta decay, Nuclear matrix element}

\date{\today}

\maketitle

%%%%%%%%%%%%%%%%%%%%%%%%%%%%%%%%%%%%%%%%%%%%%%%%%%%%%%%%%%%%%%%%%%%%%%%%%%%%

%%%%%%%%%%%%%%%%%%%%%%%%%%%%%%%%%%%%%%%%%%%%%%%%%%%%%%%%%%%%%%%%%%%%%%%%%%%%

\section{\label{sec:introduction} Introduction}
After the triumph of the neutrino oscillation and the LHC experiments in discovering two long-awaited key elements of nature, neutrino mass and mixing as well as Higgs boson, the next breakthrough of comparable 
magnitude may happen in 
neutrinoless double beta ($0\nu\beta\beta$)-decay searches. This hope is fed from both the theoretical and experimental sides. Lepton number violation (LNV) is forbidden in the Standard Model, and therefore  
observation of any LNV process would have a profound impact on particle physics and cosmology. In particular, it would prove that neutrinos are Majorana particles~\cite{Schechter:1981bd,Hirsch:2006yk}, indicate the existence of a new high-energy LNV scale and related new physics \cite{see-saw},  and provide a basis for a solution of the problem of  matter-antimatter asymmetry of the Universe via leptogenesis \cite{Fukugita:1986hr}.  Among the LNV processes, 
$0\nu\beta\beta$ decay is 
widely recognized as the most promising candidate for experimental searches.  
Another possible probe of LNV, which, as it has been recently realized, could be competitive or complementary to $0\nu\beta\beta$ decay, is the like-sign dilepton \cite{Keung:1983uu, Allanach:2009iv} searches at the LHC 
\cite{Tello:2010am,Nemevsek:2011hz,Helo:2013dla,Helo:2013ika,Helo:2013esa}. 
However,  this option still requires detailed studies to clarify its status.
On the experimental side of the $0\nu\beta\beta$ decay, one expects a significant progress in the sensitivities of near-future
experiments, stimulating the hopes for observation of this LNV process 
(for a recent review, see Ref.~\cite{Vergados:2012xy}).   

The theory of  $0\nu\beta\beta$ decay deals with three energy scales associated with rather different physics, namely,
(1) the LNV scale and underlying quark-level mechanisms of $0\nu\beta\beta$ decay, 
%: Change
(2) hadronic scale $\sim$ 1 GeV and QCD effects including nucleon form factors,  and 
(3) nuclear scale $p_{\rm F} \sim$ (100--200) MeV and nuclear structure arrangement ($p_{\rm F}$ is the nucleon Fermi momentum in a nucleus). In the literature all these 
three structure levels have been addressed from different perspectives    
(e.g., \cite{Vergados:2012xy,Deppisch:2012nb,Rodejohann:2011mu}). 

In the present paper, we revisit the mechanisms of $0\nu\beta\beta$ decay mediated by
Majorana neutrino N exchange with an arbitrary mass $m_{\rm N}$ \cite{Bamert:1994qh}. Our goal is to update and extend the analysis \cite{Benes:2005hn} of the  case with several mass eigenstates N  dominated  by ``sterile'' neutrinos $\nu_{s}$ and with an  admixture $U_{e\rm N}$ of  the active flavor 
$\nu_{e}$.
Massive neutrinos N have been considered in the literature in divers contexts 
(e.g., Ref. \cite{Abazajian:2012ys}) with the masses $m_{\rm N}$ ranging from the eV to the Planck scale. Their phenomenology has been actively studied from various perspectives including their contribution to  particle decays and production in collider experiments 
(for a recent review, see Ref.~\cite{Helo:2010cw,Atre:2009rg}). The corresponding searches for N have been carried out in various experiments \cite{Beringer:1900zz}.
An update of the previous analysis of Ref. \cite{Benes:2005hn} is needed because of the recent progress in the calculation of  the double beta-decay nuclear matrix elements  (NMEs),
which includes constraints on the nuclear Hamiltonian from the two-neutrino double-beta decay half-life 
\cite{qrpa2,anatomy}, a self-consistent description of the two-nucleon short-range correlations \cite{src09},
and the  restoration of isospin symmetry \cite{qrpawir}. 
%: Insertion-2
Our framework is given by the quasiparticle random phase approximation (QRPA).  Recently, the analysis of massive sterile neutrinos in $0\nu\beta\beta$ decay within another approach, the interacting shell model, was been carried out in
Ref. \cite{Blennow:2010th}. 
%: Insertion-2 END
There has also been significant progress 
in $0\nu\beta\beta$-decay experiments \cite{Vergados:2012xy}, especially for ${^{76}}$Ge  \cite{gelimit} and  ${^{136}}$Xe  \cite{xelimit}  
isotopes,
which allows improvements of the previous limits  in the neutrino sector.

The paper is organized as follows. 
In the next, Sec.  \ref{Formalism}, we set up the formalism underlying our analysis of the Majorana exchange mechanism of $0\nu\beta\beta$ decay. Then, we calculate the corresponding NMEs.  
Section \ref{interpolating-formula} deals with an approximate formula for the NMEs explicitly representing their dependence on $m_{\rm N}$ for arbitrary values of this parameter.  In Sec. \ref{Experimental limits}, we extract the $0\nu\beta\beta$-decay limits in the parameter plane 
$m_{\rm N}-|U_{e\rm N}|^{2}$ and compare them with other existing limits \cite{Beringer:1900zz}.  

\section{Formalism} 
\label{Formalism}
We assume that in addition to the three conventional light neutrinos there exist other
Majorana neutrino mass eigenstates $N$ of an arbitrary mass $m_{\rm N}$, dominated by the 
sterile neutrino species $\nu_{s}$ and with some admixture of the active neutrino weak 
eigenstates, $\nu_{e,\mu,\tau}$ as
\begin{eqnarray}\label{mixing}
{\rm N} = \sum\limits_{\alpha=s,e,\mu,\tau} U_{{\rm N}\alpha}\  \nu_{\alpha}.
\end{eqnarray}
The phenomenology of the intermediate mass sterile neutrinos N in various LNV processes have been actively studied in the literature (for a recent review, see Refs.~\cite{Helo:2010cw,Atre:2009rg}), and limits in 
the $|U_{\alpha \rm N}|^{2}-m_{\rm N}$-plane have been derived. It has been shown that $0\nu\beta\beta$-decay limits for $|U_{e\rm N}|^{2}-m_{\rm N}$ are the most stringent in comparison with the limits from the other LNV processes except for a narrow region of this parametric plane 
\cite{Helo:2010cw, Benes:2005hn,Mitra:2011qr}.

We study the possible contributions of  these N neutrino states to $0\nu\beta\beta$ decay via a nonzero 
admixture of a $\nu_e$ weak eigenstate. From nonobservation of this LNV process, we update 
the stringent limits on the $\nu_{\rm N}-\nu_{e}$ mixing matrix element 
$U_{e \rm N}$ in a wide region of the values of $m_{\rm N}$. We compare these limits with 
the corresponding limits derived from the searches for some other 
LNV processes. We also discuss typical uncertainties of our calculations originating 
from the models of nucleon and nuclear structure. 

The contribution of  Majorana neutrino state, N, to the $0\nu\beta\beta$-decay amplitude is described 
by the standard neutrino exchange diagram between the two
$\beta$-decaying neutrons. Assuming the dominance of this LNV mechanism,
the $0\nu\beta\beta$-decay half-life for a transition to the ground state of the final nucleus 
takes the form
\begin{equation}
[T_{1/2}^{0\nu}]^{-1} = G^{0\nu} g_{\rm A}^{4} 
\left|\sum\limits_{\rm N}\left(U^{2}_{e\rm N} m_{\rm N}\right) m_{\rm p}\, 
{M}^{\prime\, 0\nu}(m_{\rm N}, g_{\rm A}^{\rm eff}) \right|^{2},
\label{eq:1}   
\end{equation}
The proton mass is denoted by $m_{\rm p}$.  The phase-space 
factor $G^{0\nu}$ is tabulated for various $0\nu\beta\beta$-decaying nuclei in Ref. \cite{phaseint}. 
In the above formula, $g_{\rm A}$ and $g^{\rm eff}_{\rm A}$ 
stand for
the standard and ``quenched'' values of the nucleon axial-vector coupling constant, respectively. Their meanings will be discussed in what follows.
The nuclear matrix element in question, ${M}^{\prime 0\nu}$, is given by
\begin{eqnarray}
\label{eq:MnuN}
&&{M}^{\prime\, 0\nu}(m_{\rm N}, g_{\rm A}^{\rm eff})
= \frac{1}{m_{\rm p}m_{\rm e}}~
\frac{R}{2 \pi^2 g^2_A} \sum_{n} \!\! 
 \int \! d^3x \, d^3y \,  d^3p \nonumber\\
&&\times e^{i\mathbf{\rm p}\cdot (\mathbf{x}-\mathbf{y})} \frac{\bra{0^+_F} 
{J}^{\mu\dag}(\mathbf{x})
\ket{n}\bra{n}
{J}^\dag_\mu (\mathbf{y}) \ket{0^+_I}}{\sqrt{p^2+m_N^2} 
(\sqrt{p^2+m_N^2} + E_n-\frac{E_I-E_F}{2})}  \,. 
\nonumber\\
\end{eqnarray}
Here, $R$ and $m_{e}$ are the nuclear radius 
and the mass of the electron, respectively.  We use as usual  $R=r_0 A^{1/3}$ with $r_0=1.2$ fm.  
Initial and final nuclear ground states with energies 
$E_{I}$ and $E_{F}$ are denoted by $\ket{0^+_I}$ and $\ket{0^+_F}$, respectively. 
The summation runs over intermediate nuclear states $\ket{n}$ with energies $E_{n}$.
The dependence on $g_{\rm A}^{\rm eff}$ enters to 
${M}^{\prime\, 0\nu}$ through 
the weak one-body nuclear charged current  ${J}^{\dag}_\mu$ given by
\begin{eqnarray} 
\label{eq:cur1}
{J}^{0\dag}(\mathbf{r}) &=& \sum_{i=1}^A  \tau_i^+ J^0_{i} \delta
(\mathbf{r}-\mathbf{r}_i), \\
\label{eq:cur2}
\mathbf{J}^\dag(\mathbf{r}) &=& \sum_{i=1}^A \mathbf{J}_{i} 
\tau_i^+ \delta(\mathbf{r}-\mathbf{r}_i),  \,. \nonumber
\end{eqnarray}
where the sum is taken over the total number $A$ of  nucleons in a nucleus. The operators with subscript 
$i$ act only on the $i$th nucleon. The isospin rising operator
$\tau^+$ converts the neutron to a proton. 
The coordinates of beta-decaying nucleons are denoted by  ${\bf r}_{i}$. 
In the leading order of nonrelativistic 
approximation, one has
\begin{eqnarray}
\label{NRL-J}
J^{0 \dagger} &=& g_{\rm V}(p^2), \nonumber\\
{\mathbf J}^{\dagger} &=& - g_{\rm A}(p^2) \boldsymbol{\sigma}
+ g_{\rm P}(p^2)  \frac{  \mathbf{p}(\boldsymbol{\sigma}\cdot\mathbf{p})}{2 m}
\nonumber\\
&& - i \left(g_{\rm V}(p^2) + g_{\rm M}(p^2)\right)  
\frac{\boldsymbol{\sigma}\times\mathbf{p}}{2 m}.
\end{eqnarray}
Here, $\mathbf{p}=\mathbf{p}_n-\mathbf{p}_p$, with $\mathbf{p}_n$ and $\mathbf{p}_p$ being the initial neutron and the final proton 3-momenta, respectively.
For the nucleon electroweak form factors, we use the standard parametrization,
\begin{eqnarray}\label{ffn-1}
g_{\rm V}({ p}^{~2}) &=& \left(1+\frac{{ p}^{~2}}{{M^2_V}}\right)^{-2},\ \ \  \
g_{\rm A}({ p}^{~2}) = g^{\rm eff}_{\rm A}\left(1+\frac{{ p}^{~2}}{{M^2_A}}\right)^{-2}, 
\nonumber\\ 
g_{\rm M}({ p}^{~2}) &=& (\mu_p-\mu_n)~ g_{\rm V}({ p}^{~2}),\nonumber\\  
g_{\rm P}({ p}^{~2}) &=& 2 m_{\rm p} g_{\rm A}({ p}^{~2})({ p}^{~2} + m^2_\pi)^{-1},
\end{eqnarray}
where $(\mu_p-\mu_n) = 3.70$, $M_V = 850$ MeV, $M_A = 1086$ MeV,
and $m_\pi$ is the pion mass.  For the induced pseudoscalar form factor $g_{\rm P}(p^{~2})$, 
the standard Goldberger--Treiman PCAC relation is used.

The value of the nucleon axial-vector coupling constant in vacuum is $g_A=1.269$. 
In the nuclear medium this constant is expected to be renormalized to some smaller, the so-called quenched, value 
$g^{\rm eff}_A$ \cite{quenching}. 
This is motivated, in particular, by the fact that 
the calculated values of the strength of the Gamow--Teller $\beta$-decay transitions to individual 
final states are significantly larger than the experimentally measured ones. 
Theoretically, the Gamow--Teller strength is a monotonically increasing function of $g_{\rm A}$. Therefore,  this discrepancy with experiment can be rectified by a proper adjustment of  $g_{\rm A}$ to some  smaller quenched value  $g^{\rm eff}_{\rm A}$. It was shown in Refs. \cite{qrpa2,qrpa2}, that this value is compatible with
the quark axial-vector coupling $g^{\rm eft}_{\rm} = g_{\rm A}^{\rm quark}=1$. In some recent works 
$g_{\rm A}^{\rm eff}< 1$ has been advocated
\cite{fogliga,Barea:2013wga}. 
In our opinion, this sort of strong quenching still requires a more firm justification.  
Therefore, in our analysis, we consider the following two options:
\begin{eqnarray}\label{gA}
g^{\rm eff}_A &=& g_{\rm A}=1.269  \hspace{30mm} \mbox{\cite{Beringer:1900zz}},  
\\
\label{gAQ}
g^{\rm eff}_A  &=& g_{\rm A}^{\rm quark} =1  \hspace{31mm}  \mbox{\cite{qrpa2,qrpa2}}.   
\end{eqnarray}
%
%%%%%%%%%%%%%%%%%%%%%%%%%%%%%%%%%%%
%
%\vspace*{-5mm}
\begin{figure}[!t]
\vspace*{-5mm}
\begin{center}
    \includegraphics[height = 7cm]{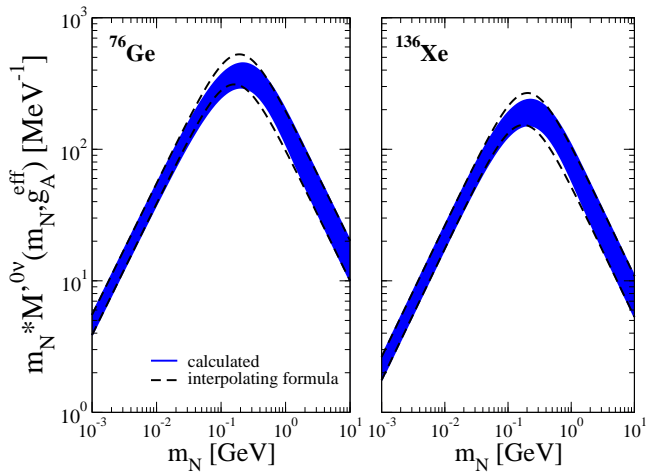}
  \end{center}
\vspace*{-5mm}
\caption{The product $m_{\rm N} M^{0\nu}(m_{\rm N})$ vs mass of heavy neutrino 
$m_{\rm N}$ for ${}^{76}$Ge and ${}^{136}$Xe 
within QRPA with partial restoration of isospin symmetry \cite{qrpawir}. The filled  blue area represents the uncertainty associated 
with the choice of the NN potential  (CD--Bonn and Argonne potentials) and the value of the nucleon axial-vector constant 
($g_A^{\rm eff}=1.0$ and  1.269). The dashed lines and the area between them correspond to results obtained with
the approximate formula in Eq. (\ref{interpol})}
\label{fig.1}
\end{figure}
%
%%%%%%%%%%%%%%%%%%%%%%%%%%%%%%%%%%%

We calculated the NME defined in Eq. (\ref{eq:MnuN}) within 
the QRPA with partial restoration of isospin symmetry \cite{qrpawir}.
Two different types of NN potentials (CD--Bonn and Argonne) as well as unquenched and 
quenched values of the nucleon axial-vector coupling in Eqs. (\ref{gA}) and (\ref{gAQ}) were considered.
The results for 
the particular cases of ${}^{76}$Ge and ${}^{136}$Xe we show in \mbox{Fig. \ref{fig.1}}.  
The widths of the blue bands illustrate the typical uncertainties of our approach related to the choice of the NN potential and the value of $g_A^{\rm eff}$. 

\section{``Interpolating''  formula}
\label{interpolating-formula}
We have also carried out the calculations of the NME in Eq. (\ref{eq:MnuN}) for the two conventional limiting cases: the light 
$m_{\rm N}\ll p_{\rm F}$ and the heavy  $m_{\rm N} \gg p_{\rm F}$ Majorana neutrino exchange mechanisms,
where $p_{\rm F}\sim$ 200 MeV is the characteristic momentum transferred via the virtual neutrino, which is of the order of the mean nucleon momentum of Fermi motion in a nucleus.
For these limiting cases the half-life formula (\ref{eq:1}) is reduced to
\begin{eqnarray}
&&[T_{1/2}^{0\nu}]^{-1} = G^{0\nu} g_{\rm A}^{4}\times \nonumber\\  
%\begin{array}
%\end{array}
&&\times \left\{\begin{array}{ll}
    \left|\frac{\langle m_{\nu}\rangle}{m_{\rm e}}\right|^{2} 
     \left|{M}^{\prime 0\nu}_{\nu}(g^{\rm eff}_{\rm A})\right|^{2},  
& \mbox{for}\  m_{\rm N} \ll p_{\rm F},    \\[3mm]
\left|\langle\frac{1}{m_{\rm N}}\rangle m_{\rm p}\right|^{2} 
   \left|{M}^{\prime 0\nu}_{\rm N}(g^{\rm eff}_{\rm A})\right|^{2}, 
    &  \mbox{for}\  m_{\rm N} \gg p_{\rm F},  
\end{array}\right.
\label{LightHeavy}   
\end{eqnarray}
with 
\begin{eqnarray}\label{meanMass}
\langle m_{\nu}\rangle = \sum\limits_{\rm N} U_{e\rm N}^{2} m_{\rm N},\ \ \ \ 
%\label{meanMass-2}
\left\langle\frac{1}{m_{\rm N}}\right\rangle =  \sum\limits_{\rm N} \frac{U_{e\rm N}^{2}}{m_{\rm N}}.
\end{eqnarray}
Here, the NMEs ${M}^{\prime 0\nu}_{\nu}, {M}^{\prime 0\nu}_{N}$ are derived from the NME 
${M}^{\prime 0\nu}$ in Eq. (\ref{eq:MnuN}) in the following way:
\begin{eqnarray}\label{lim-rel-1}
{M}^{\prime 0\nu} (m_{\rm N}\rightarrow 0, g^{\rm eff}_{\rm A})
&=&  \frac{1}{m_{\rm p} m_{\rm e}} {M}^{\prime 0\nu}_{\nu}(g^{\rm eff}_{\rm A}) 
,\\
\label{lim-rel-2}
{M}^{\prime 0\nu} (m_{\rm N}\rightarrow \infty, g^{\rm eff}_{\rm A})
 & = & \frac{1}{m_{\rm N}^{2}} {M}^{\prime 0\nu}_{\rm N}(g^{\rm eff}_{\rm A}).
\end{eqnarray}
The values of ${M}^{\prime 0\nu}_{\nu}(g^{\rm eff}_{\rm A})$ and  
${M}^{\prime 0\nu}_{\rm N}(g^{\rm eff}_{\rm A})$  calculated in the QRPA with partial restoration of isospin symmetry \cite{qrpawir} for all the experimentally interesting isotopes  are given in Table \ref{table.1} (for more details of the formalism, see Refs.~\cite{qrpa2,anatomy,src09}).  
These NMEs can be used for the analysis of  the light and the heavy Majorana exchange mechanisms of 
$0\nu\beta\beta$ decay on the basis of Eqs. (\ref{LightHeavy}).

%%%%%%%%%%%%%%%%%%%%%%%%%%%%%%%%%%%%%%%%%%%%%%%%%
%
\begin{table*}[htb] 
\caption{
The values of the nuclear matrix elements for 
the light and  heavy neutrino mass mechanisms defined
in Eqs. (\ref{lim-rel-1}) and (\ref{lim-rel-2}) and the parameters $\langle p^{2}\rangle$ and ${\cal A}$
of  the interpolating formula specified in Eqs. (\ref{interpol})--(\ref{coeff-2}). The  
calculations have been carried out within the QRPA with partial restoration of isospin symmetry \cite{qrpawir}.
Two different types of NN potential (CD--Bonn and Argonne) as well as quenched ($g_A =1.00$) 
and unquenched ($g_A =1.269$) values of the nucleon axial-vector constant have been considered.
%The notation is as follows: 
The cases presented are
a) Argonne potential, $g_A =1.00$; 
b) Argonne, $g_A =1.269$; c) CD--Bonn, $g_A =1.00$; and 
\mbox{d) CD--Bonn, $g_A =1.269$}.  
}
\label{table.1}    
\centering 
\renewcommand{\arraystretch}{1.1}  
%\addtolength{\tabcolsep}{-1pt} 
%\scriptsize 
\begin{tabular}{lcccccccccccccccccccc}
\hline \hline 
nucleus & &  \multicolumn{4}{c}{${M'}^{0\nu}_\nu$} & & \multicolumn{4}{c}{${M'}^{0\nu}_N$} 
& & \multicolumn{4}{c}{$\sqrt{\langle p^{2}\rangle}$ [MeV]} 
& & \multicolumn{4}{c}{${\cal A}$ [$10^{-10} \mbox{yrs}^{-1}$]} \\
\cline{3-6}  \cline{8-11} \cline{13-16}  \cline{18-21}    
       & & a & b & c & d & & a & b & c & d & & a & b & c & d & & a & b & c & d \\ \hline
  ${^{48}Ca}$   & & 0.463 & 0.541 & 0.503 & 0.594 & &  29.0 & 40.3 & 49.0 & 66.3 & & 173.0 & 189.0& 216.0 & 231.0 & &
 0.541  & 1.05  &  1.55 &  2.83 \\
  ${^{76}Ge}$   & & 3.886 & 5.157 & 4.211 & 5.571 & &  204.0 & 287.0 & 316.0 & 433.0 & & 159.0 & 163.0 & 190.0 & 193.0 & &
 2.55  & 5.05  &   6.12 &   11.5 \\
  ${^{82}Se}$   & & 3.460 & 4.642 & 3.746 & 5.018 & &  186.0 & 262.0 & 287.0 & 394.0 & & 161.0 & 165.0 & 192.0 & 194.0 & & 
 9.12 & 18.1 &  21.7 & 40.9 \\
  ${^{96}Zr}$   & & 2.154 & 2.717 & 2.341 & 2.957 & &  132.0 & 184.0 & 202.0 & 276.0 & & 171.0 & 180.0 & 203.0 & 212.0 & & 
 9.30 & 18.1 &  21.8 & 40.7 \\
  ${^{100}Mo}$  & & 4.185 & 5.402 & 4.525 & 5.850 & &  244.0 & 342.0 & 371.0 & 508.0 & & 167.0 & 174.0 & 198.0 & 204.0 & &
 24.6 & 48.3 &  56.8 & 107. \\
  ${^{110}Pd}$  & & 4.485 & 5.762 & 4.856 & 6.255 & &  238.0 & 333.0 & 360.0 & 492.0 & & 160.0 & 166.0 & 189.0 & 194.0 & &
 7.07 & 13.8 &  16.2 & 30.2 \\
  ${^{116}Cd}$  & & 3.086 & 4.040 & 3.308 & 4.343 & &  150.0 & 209.0 & 222.0 & 302.0 & & 153.0 & 157.0 & 179.0 & 183.0 & &
 9.74 & 18.9 &  21.3 & 39.5 \\
  ${^{124}Sn}$  & & 2.797 & 2.558 & 3.079 & 2.913 & &  146.0 & 184.0 & 224.0 & 279.0 & & 158.0 & 186.0 & 187.0 & 214.0 & & 
 5.00 & 7.94 &  11.8 & 18.2 \\
  ${^{128}Te}$  & & 3.445 & 4.563 & 3.828 & 5.084 & &  215.0 & 302.0 & 331.0 & 454.0 & & 173.0 & 178.0 & 204.0 & 207.0 & & 
 0.705 &  1.39  & 1.67 &  3.14 \\
  ${^{130}Te}$  & & 2.945 & 3.888 & 3.297 & 4.373 & &  189.0 & 264.0 & 292.0 & 400.0 & & 175.0 & 180.0 & 206.0 & 209.0 & & 
 13.2 & 25.7 & 31.4 & 59.0 \\
  ${^{136}Xe}$  & & 1.643 & 2.177 & 1.847 & 2.460 & &  108.0 & 152.0 & 166.0 & 228.0 & & 178.0 & 183.0 & 208.0 & 211.0 & &
 4.41 &  8.74 & 10.4 & 19.7 \\
\hline \hline  
\end{tabular}  
\end{table*}    
%%%%%%%%%%%%%%%%%%%%%%%%%%%%%%%%%%%%%%%%%%%%%%%%%%%%%%%%%%%%%%

What we would like to highlight here is that these limiting-case NMEs also allow one to approximate the NME or half-life for arbitrary $m_{\rm N}$
with the aid of a useful ``interpolating formula'' proposed in Ref. \cite{Kovalenko:2009td} and used in the literature (e.g., \cite{Helo:2010cw,Mitra:2011qr,*Abada:2014nwa}) in analysis of $0\nu\beta\beta$ decay. For the half-life, it reads
\begin{eqnarray}\label{interpol}
[T_{1/2}^{0\nu}]^{-1} =  {\cal A} \cdot
\left|m_{\rm p} \sum\limits_{\rm N}U^{2}_{e\rm N} \frac{m_{\rm N}} {\langle p^{2}\rangle + m_{\rm N}^{2}} \right|^{2},
\end{eqnarray}
where 
\begin{eqnarray}
\label{coeff-1}
{\cal A} \ \ &=& G^{0\nu} g_{\rm A}^{4}  \left|M^{\prime 0\nu}_{\rm N}(g^{\rm eff}_{\rm A})  \right|^{2}, \\ 
\label{coeff-2}
\langle p^{2}\rangle &=& m_{\rm p} m_{\rm e} \left|\frac{M^{\prime 0\nu}_{\rm N}(g^{\rm eff}_{\rm A}) }
{M^{\prime 0\nu}_{\nu}(g^{\rm eff}_{\rm A}) } \right|^{2}
\end{eqnarray}
with the values of the matrix elements $M^{\prime 0\nu}_{\nu}, M^{\prime 0\nu}_{\rm N}$  and parameters 
$\langle p^{2}\rangle$ and ${\cal A}$ given for various isotopes in Table \ref{table.1}. 
To estimate the accuracy of the approximate formula (\ref{interpol}) we compare it with 
the ``exact'' QRPA results in Fig.~\ref{fig.1} for ${}^{76}$Ge and ${}^{136}$Xe, where the dotted curves correspond to the interpolating formula (\ref{interpol}). As seen, it is a rather good approximation of the exact QRPA result except for the transition region in which the accuracy is about 
20\% -- 25\%. 

The clear advantage of the formula (\ref{interpol}) is that it shows explicitly the $m_{\rm N}$ dependence of the $0\nu\beta\beta$ amplitude or the half-life. Therefore, it can be conveniently used for an analysis of any contents of the neutrino sector without engaging the sophisticated machinery of the nuclear structure calculations. Also any upgrade of nuclear structure approaches typically bringing out asymptotical NMEs for $m_{\rm N}\ll p_{\rm F}$ and $m_{\rm N}\gg p_{\rm F}$ allows one to immediately reconstruct with a good accuracy updated NMEs  for arbitrary $m_{\rm N}$.   

For completeness, let us give the $0\nu\beta\beta$-decay half-life formula for a generic neutrino spectrum,
which incorporates a popular scenario $\nu$MSM \cite{Asaka:2005pn,Asaka:2005an}, offering a solution of the dark matter DM and baryon asymmetry (BAU) problems via massive Majorana neutrinos. 
%: Insertion-1
%Implications of  the $\nu$MSM for $0\nu\beta\beta$-decay have been considered in 
%
In Refs. \cite{Bezrukov:2005mx,*Asaka:2011pb,*Merle:2013ibc}, $0\nu\beta\beta$ decay has been considered within the $\nu$MSM  employing  certain approximations in order to estimate $0\nu\beta\beta$-decay half-life.  
We note that our Eq. (\ref{interpol}) offers a suitable and systematic tool for this purpose especially when both small and large values of $m_{N}$ are involved. 
%: Insertion-1 END

Let the neutrino spectrum contain  (i) three light neutrinos $\nu_{k=1,2,3}$ with the masses $m_{\nu (k)}\ll p_{\rm F}\sim$200 MeV dominated by $\nu_{e,\mu,\tau}$, (ii) a number of the DM candidate neutrinos 
$\nu^{DM}_{i}$ with the masses $m^{DM}_{i}$ at the keV scale, (iii) a number of heavy neutrinos $N$ with the masses $m_{\rm N}\gg p_{\rm F}$, plus (iv)  several intermediate mass $m_{h}$ neutrinos $h$ among which there could be a pair highly degenerate in mass needed for 
the generation of the BAU via leptogenesis \cite{Asaka:2005an}.  In this case, the interpolating formula  (\ref{interpol}) allows us to write down for the half-life of any $0\nu\beta\beta$-decaying isotope
\begin{eqnarray}\label{Example}
&&[T_{1/2}^{0\nu}]^{-1} =  {\cal A} 
\left|\frac{m_{\rm p}}{\langle p^{2}\rangle}  
\sum\limits_{k=1}^{3}U^{2}_{ek}m_{k}  +
\frac{m_{\rm p}}{\langle p^{2}\rangle}  \sum\limits_{i}\left(U^{DM}_{ei}\right)^{2}m^{DM}_{i}  
\right.\nonumber\\
&& \left.~~~~~~~~~~~~~~~~~~ + m_{\rm p}
\sum\limits_{\rm N}\frac{U^{2}_{e\rm N} }{ m_{\rm N}} + m_{\rm p}
\sum\limits_{h}\frac{U^{2}_{eh}  m_{h}} {\langle p^{2}\rangle + m_{h}^{2}} \right|^{2}. 
\end{eqnarray}
Here, bedause of typically very small mixing between the light and massive neutrino mass eigensates 
\mbox{$|U^{DM}_{ei}|, |U_{e\rm N}|,$ and $|U_{eh}|\ll |U_{ek}|$} the mixing matrix of the light neutrinos $\nu_{k}$ to a good accuracy 
can be identified with the element of the PMNS mixing matrix 
\mbox{$U_{ek} \approx U^{PMNS}_{ek}$}.
%
%%%%%%%%%%%%%%%%%%%%%%%%%%%%%%%%%%%
\begin{figure}[!t]
\vspace*{-7mm}
  \begin{center}
    \includegraphics[height = 7cm]{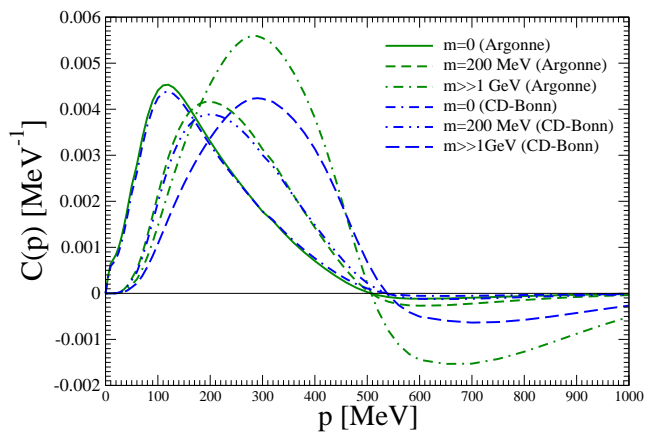}
  \end{center}
\vspace*{-5mm}
  \caption{The normalized momentum transfer $p$ distribution  $C(p)$ \cite{qrpawir} of the 
virtual neutrino characterizing its contribution to the nuclear matrix element (\ref{eq:MnuN}) 
in the function of  $p$.}
\vspace*{-5mm}
\label{fig.2}
\end{figure}
%%%%%%%%%%%%%%%%%%%%%%%%%%%%%%%%%%%

Finally, the following observation might be of interest.  Note that the parameter $\langle p^{2}\rangle $ with the typical value 
\mbox{$\sim$ (200 MeV)${}^{2}$ }can be interpreted as the mean Fermi momentum of nucleons  $p_{\rm F}$ in a nucleus. 
This is suggested by the structure of  the NME in Eq. (\ref{eq:MnuN}). In fact, 
we can schematically write for the $m_{\rm N}$ dependence 
\begin{eqnarray}\label{justif}
%
%&&\langle 0^{+}_{F}| H(r_{kl}, m_N) | 0^{+}_{I}\rangle \nonumber\\
M^{\prime 0\nu}(m_{\rm N}) &\simeq& {\rm const}\cdot \int_{0}^{\infty} 
%\langle 0^{+}_{F}|  j_{0,2}(p r_{kl}) | 0^{+}_{I}\rangle
\frac{\ h(p^2)~p^2  ~d{p}}
{\sqrt{p^2+m^2_N} (\sqrt{p^2+m^2_N}
+ \overline{E}_n)} \nonumber\\
&\simeq& {\rm const}\cdot  \frac{1}{\overline{p^{2}} + m_{\rm N}^{2}} \equiv 
{\rm const} \cdot \frac{1}{\langle p^{2}\rangle + m_{\rm N}^{2}}.
%h_{K}(p^2)~q
%
\end{eqnarray}
Here, $\overline{E}_n = E_n-(E_I-E_F)/2 \sim$ 10 MeV is a small value in comparison with the so-defined 
mean neutrino momentum $\overline{p^{2}}$, taking into account the smearing effect of 
the nucleon form factors and the nuclear wave function codified in the $h(p^2)$ factor (for definitions, see Ref. \cite{anatomy}).
In the last step in Eq. (\ref{justif}), we identified $\overline{p^{2}}$ with the parameter $\langle p^{2}\rangle$
in Eq. (\ref{interpol}) as suggested by the comparison of Eq. (\ref{interpol}) with Eq. (\ref{justif}).
Kinematically, the mean momentum transfer such us $\sqrt{\langle p^{2}\rangle}$ is expected to be of the order of the mean nucleon Fermi momentum $p_{\rm F}$ in a nucleus. 

Although $\langle p^{2}\rangle$ is just a parameter of the parametrization (\ref{interpol}) tabulated in Table \ref{table.1}, its  rather small variation over the isotopes supports the above physical interpretation. On top of that, we show in Fig. \ref{fig.2} the normalized momentum transfer distribution $C(p)$ 
defined in Ref. \cite{qrpawir}. It characterizes  the contribution of the momentum $p$ to the NME for several values of $m_{\rm N}$ and two options for the NN potential. As seen from Fig. \ref{fig.2} for the intermediate mass $m_{\rm N}=$200 MeV corresponding to the transition region of the interpolating formula 
in Eq. (\ref{interpol}),  the NME is dominated by the mean value of the virtual neutrino momentum $p\approx$ 200 MeV. This fact again indicates that the parameter $\sqrt{\langle p^{2}\rangle}$ is correlated with the mean momentum transfer and, consequently, with $p_{\rm F}$.
The above-given interpretation could be useful for gross estimates analyzing systems for which the NMEs are unavailable.

\section{Experimental limits}
\label{Experimental limits}
\begin{figure}[!t]
\vspace{-10mm}
\hspace*{-10mm}
 % \begin{center}
%\includegraphics[width=10.6cm]{U2-mN-1.pdf}
   \includegraphics[width=10.6cm]{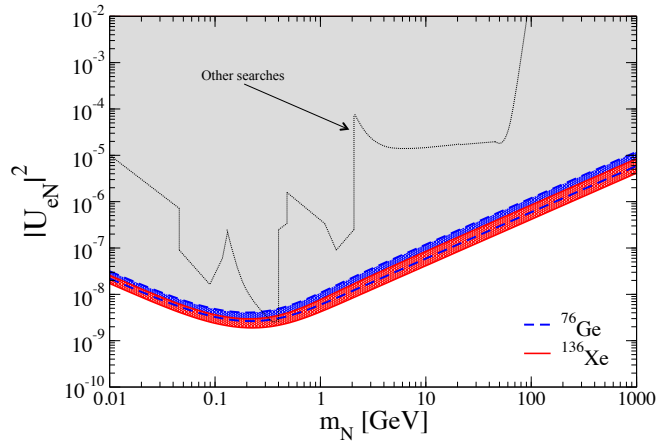}
 % \end{center}
\vspace{-12mm}
  \caption{Exclusion plots in the $|U_{e\rm N}|^2-m_{\rm N}$ plane. 
The band restricted by blue dashed lines (red solid lines) is the lower limit from the experimental searches 
for  $0\nu\beta\beta$ decay of ${}^{76}$Ge (${}^{136}$Xe). The weakest (strongest) limit is obtained for $M^{0\nu}(m_{\rm N})$ calculated with Argonne potential (CD--Bonn potential) and assuming $g_A=1.00$ ($g_A=1.269$).
The thin dotted line other searches shows a region excluded from the various laboratory searches for massive neutrinos
\cite{Beringer:1900zz,Atre:2009rg}.
\label{fig.3}}
\end{figure}

Having the nuclear matrix element ${M'}^{0\nu}(m_{\rm N})$ calculated, we can derive 
the $0\nu\beta\beta$-decay limits on the mass $m_{\rm N}$ of the N neutrino 
and its mixing $U_{e \rm N}$ with the $\nu_e$ neutrino weak eigenstate. Here, we assume no significant cancellation between different terms in Eq. (\ref{eq:1}) or (\ref{interpol}). In other words, we consider only one term in Eqs. (\ref{eq:1}) and (\ref{interpol}).
Applying the presently best lower bounds on the $0\nu\beta\beta$-decay half-life of $^{76}$Ge 
(combined GERDA + Heidelberg--Moscow) \cite{gelimit} 
and $^{136}$Xe (combined EXO+KamlandZEN) \cite{xelimit},
\begin{eqnarray}\label{HD-M}
T_{1/2}^{0\nu}({^{76}Ge})
&\geq& T_{1/2}^{0\nu-exp}({^{76}Ge}) \ \, = 3.0~ 10^{25}~ \mbox{yrs},\\
T_{1/2}^{0\nu}({^{136}Xe})
&\geq& T_{1/2}^{0\nu-exp}({^{136}Xe}) = 3.4~ 10^{25}~ \mbox{yrs},\nonumber
\end{eqnarray}
we derived from Eq. (\ref{eq:1}) the $|U_{e \rm N}|^2-m_{\rm N}$ exclusion plot shown 
in Fig.~\ref{fig.3}. 
Alternatively, as we demonstrated in Sec. \ref{interpolating-formula}, the same could be done on 
the basis of the interpolating formula in Eq. (\ref {interpol}) without visible changes in Fig.~\ref{fig.3}. 
%
%%%%%%%%%%%%%%%%%%%%%%%%%%%%%%%%%%%
%%%%%%%%%%%%%%%%%%%%%%%%%%%%%%%%%%%

In Fig.~\ref{fig.3}
we also show typical domains excluded by some other experiments summarized in 
Refs.~\cite{Beringer:1900zz,Atre:2009rg}. 
These domains are just indicative, because most of the previous bounds were obtained for 
some fixed values of $m_{\rm N}$. 
For convenience, we interpolated this set of experimental points by continuous curves
in different intervals of $m_{\rm N}$. 
As seen from Fig.~\ref{fig.3},  the $0\nu\beta\beta$-decay limits 
%cover much wider region of masses $m_N$ and 
exclude the parts of the $|U_{e \rm N}|^2-m_{\rm N}$ parameter space previously unconstrained by the laboratory 
experiments except for a very small interval $m_{\rm N}=$300--400 MeV. 
%: Insertion-3
\footnote{Note that our exclusion plot in Fig.~\ref{fig.3}  is given for $m_{N}\geq 10$ MeV, where other constraints  
~\cite{Beringer:1900zz,Atre:2009rg} for comparatively heavy N are located. Obviously, it can be extrapolated both in $m_{N}\rightarrow 0$ and $m_{N}\rightarrow \infty$ directions since our approach is valid for arbitrary $m_{N}$. Outside the region of  $m_{N}$ in Fig.~\ref{fig.3} our curve is given with a good accuracy by the second and the third terms of Eq. (\ref{Example}).}
%: Insertion-3 END
%
However, the following comment here is in order. The constraints listed in Refs. \cite{Beringer:1900zz} are based on the 
searches for peaks in differential rates of various processes and the direct production of N states 
followed by their decays in a detector. In Refs. \cite{Helo:2010cw,SP}, it was pointed out that in 
this case the results of data analysis depend on the
total decay width of N, including the neutral current decay channels. The latter have not been properly 
taken into account in the derivation of the mentioned experimental constraints. 
However, the neutral current N-decay channels introduce the dependence of the final results on all the mixing matrix elements $U_{e \rm N}, U_{\mu \rm N}$, and $U_{\tau \rm N}$. In this situation, one cannot extract individual limits for these matrix elements without some additional assumptions, introducing a significant uncertainty. 
In contrast, our $0\nu\beta\beta$-decay limits involve only the $U_{e \rm N}$ mixing matrix element and therefore are free of the mentioned uncertainty. 
This is because in $0\nu\beta\beta$ decay intermediate Majorana neutrinos are always off-mass-shell states and their decay widths are irrelevant.  
On the other hand, the above-derived $0\nu\beta\beta$-decay constraints may be significantly weakened in the presence of the $CP$ Majorana phases $\alpha^{CP}\neq 2 \pi n$, for an integer $n$.  This is because, in that case, in Eqs. (\ref{eq:1}), (\ref{interpol}),  and (\ref{Example}), a cancellation between different terms may happen.  

\section{Summary}
\label{Summary}
We updated the $0\nu\beta\beta$-decay limits in the plane $|U_{e\rm N}|^{2}-m_{\rm N}$ for the updated nuclear matrix elements \cite{qrpawir} and experimental data (\ref{HD-M}). Our limits are shown in 
Fig. \ref{fig.3}.  We studied some uncertainties endemic to the nuclear structure calculations in general and for the QRPA in particular. These are the choice of the NN-potential and the value of the nucleon 
axial-vector coupling $g^{\rm eff}_{\rm A}$ in nuclear matter. In Fig. \ref{fig.3}, we compared the $0\nu\beta\beta$-decay limits with the corresponding limits from other searches and showed that the former confidently override the latter for all $m_{\rm N}$ values except for a narrow interval around $\sim$300 MeV at which certain improvement of the $0\nu\beta\beta$-decay limits is needed. We also commented on the reliability of both the experimental results shown in Fig. \ref{fig.3} as ``other searches'' and the $0\nu\beta\beta$-decay limits themselves disclosing some assumptions incorporated in their derivation. 

We analyzed the interpolating formula, Eq. (\ref{interpol}), from the viewpoint  of its accuracy and usefulness in phenomenological analysis of neutrino models in the part of their predictions for 
$0\nu\beta\beta$ decay. This formula allows one to easily update $0\nu\beta\beta$-decay limits for 
$|U_{\rm eN}|^{2}-m_{\rm N}$ once either new experimental data for the $0\nu\beta\beta$-decay half-life or updated NMEs for the light and heavy Majorana exchange mechanisms are released. As an application of this formula we gave an approximate representation of the $0\nu\beta\beta$-decay half-life in Eq. (\ref{Example}) for the neutrino spectrum of the presently popular $\nu$MSM scenario \cite{Asaka:2005pn,Asaka:2005an}.

\begin{acknowledgments}

F. \v S.\ acknowledges the support by the VEGA Grant agency of the Slovak Republic under the contract No.\ 1/0876/12 
and by the Ministry of Education, Youth and Sports of the Czech Republic under contract LM2011027.
The work was partially supported by the PIIC fellowship (M.G.) from 
Universidad T\'ecnica Federico Santa Mar\'ia (UTFSM)  and the research grant (S.K.) from the UTFSM.
\end{acknowledgments}

%\bibliography{references_0nubb}
%\bibliography{references_0nubb_M}
%\bibliographystyle{h-physrev5}

\end{document}